# How Saturn could create rings by itself. The third force of diamagnetic expulsion and the mechanism of the magnetic anisotropic accretion of the origin of Saturn's rings

**Vladimir V. Tchernyi\* and Sergey V. Kapranov †**

\* Modern Science Institute, SAIBR. Osennii blvd, 20-2-702, Moscow, 121614, Russia.
  Email: chernyv@bk.ru
† Moscow representative office of A.O. Kovalevsky Institute of Biological of the Southern Seas, Russian Academy of Sciences, Moscow 119991, Russia. Email: sergey.v.kapranov@yandex.ru

## Abstract

Here we demonstrate how Saturn could create its rings by itself, with the additional action of its axisymmetric magnetic field once appeared. A new mechanism for the origin of Saturn's rings is proposed for the first time. It includes the appearance of an additional third force - the force of diamagnetic expulsion of a diamagnetic ice particle after the emergence of the magnetic field of Saturn and the mechanism of magnetic anisotropic accretion. This force is acting together with force of gravity and centrifugal one on the particles within protoplanetary cloud. The model is allowed understand many of unexplainable phenomena observed in the rings. Cassini found 93% of ice in the particles of rings. There is no complete knowledge of ice in space. We know 17 types of ice on Earth. Ice of the XI type has stable parameters at the temperature of the rings and it is diamagnetic. The presence of Saturn's magnetic field and the low temperatures near the rings lead us to the idea of diamagnetism of the particles of rings. The rings could have originated from ice particles moving under the influence of centrifugal force and force of gravity in chaotic orbits around Saturn within protoplanetary cloud after the planet's magnetic field was emerged. After appearance of the force of diamagnetic expulsion of ice particles, all their chaotic orbits start shifting to the magnetic equator plane, where the minimum of magnetic energy of the particles is observed. Every particle on the magnetic equator comes to a stable position, and it prevents its horizontal and vertical shift. The particles are trapped within three-dimensional magnetic well. Other theories of the rings origin are considered earlier by other authors may contribute some features to the final picture of the rings. In the proposed theory, the known knowledge about their contribution to the origin of the rings of Saturn is not denied, but complemented with appearance of the third force of diamagnetic expulsion and additional mechanism of magnetic anisotropic accretion.

**Keywords:** Saturn rings origin, magnetic anisotropic accretion, ice of Saturn's rings, diamagnetism of Saturn's ice particles, space diamagnetism, orthorhombic ice XI.

1. **About history of appearance of a new concept of magnetic anisotropic accretion**

Questions about the origin, evolution, and age of Saturn's rings have remained unanswered since G. Galileo first saw them in 1610. J. Maxwell proved that rings consist of particles (1856), G. Kuiper predicted that particles consist of ice (1947), and the Cassini probe (2004-2017) found that ring particles consist of 93% ice [1]-[3].

The ideas of the well-known explanation of the origin of Saturn's rings are based on the gravitational defragmentation of a massive body that approached Saturn or comet tidal disruption, etc. [4]-[8]. What is missing in these scenarios is how the sombrero disk of rings is obtained, so well constructed with separated particles and a fine ring structure with strong flatness. This poorly explains the horizontal and vertical stability of the particles in the rings, the stability of the rings themselves, the appearance of spokes in the B ring, and electromagnetic phenomena in the rings. There is also no answer to the question of why planets located inside the asteroid belt do not have rings.

If we compare the ratio of the thickness of the rings to their diameter, with the thickness of a sheet of A4 paper to its length, then the relative thickness of the disk of the rings will be a thousand times smaller. It is amazing that such a thin film of ice particles of huge diameter hangs in outer space. The question arises whether all possible interactions in the Saturn-ice particle system of the protoplanetary cloud were taken into account earlier and whether the possible role of other interactions in the origin of the rings, which have not yet been described, is not overlooked. Therefore, we are trying to fill this gap in the knowledge of ring particles and try to understand how the diamagnetism of ice particles affects the formation of rings.

2. **The concept of magnetic anisotropic accretion**

The theoretical concept is that after the appearance of the magnetic field of Saturn and the appearance of the diamagnetic expulsion force for ice particles, all the chaotic orbits of the particles inside the protoplanetary cloud began to shift to the plane of the magnetic equator. As a result, the protoplanetary cloud surrounding Saturn collapses into a disk with orbits of particles at the planet's magnetic equator. And as a result, all the particles form a disk system of rings. The gravitational force in the orbit of the particle is balanced by the centrifugal force and the force of diamagnetic expulsion.

The problem to be solved is how, after the interaction of the magnetic field of Saturn with the diamagnetic ice particles of the protoplanetary cloud of Saturn, all the orbits of the ice particles can come to the plane of the magnetic equator and create a system of rings. All the ice particles at the end of the motion must be trapped inside a three-dimensional magnetic well in the plane of Saturn's magnetic equator.

3. **Solution of the problem of contribution of magnetic anisotropic accretion to the origin of rings**

We need to demonstrate how protoplanetary cloud can collapse into a disk of rings (Fig. 1).

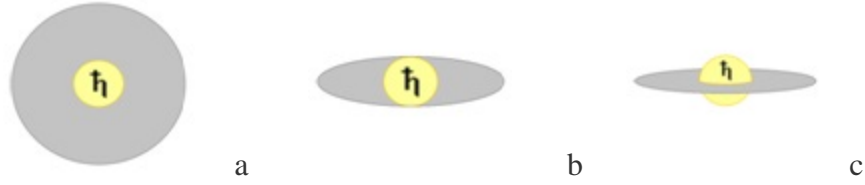

Fig. 1. Transformation of Saturn's protoplanetary cloud into a disk of rings after appearance of Saturn's magnetic field and interaction of it with the iced particles: from (a) >> (b) >> to (c)

To do this we need to solve the problem of how after the emergence of the magnetic field of Saturn all chaotic orbits of diamagnetic ice particles of the protoplanetary cloud of Saturn can move in the plane of the magnetic equator and create a system of rings with separated particles and dividing thin structure of rings. It is important to note that for Saturn we have a spherically symmetric gravitational field and an axisymmetric magnetic field.

We use V. Safronov's theory of the small nebula [9]. He received the G. Kuiper Prize in 1990.

The mathematical solution of the problem is based on the fundamental theory presented by V. Tchernyi and S. Kapranov [10, 11].

First, we solve the problem of a single diamagnetic spherical particle located in the external gravitational field and magnetic field of Saturn. The solution of the problem of the motion of diamagnetic particles after the emerging of the magnetic field of Saturn leads us to the equation for the azimuthal angle of the particles:

$$\ddot{\theta}_{S-p} + \dot{\theta}_{S-p} \cot\theta_{S-p} = \frac{GM_S}{r_{S-p}^3}\cot\theta_{S-p} + \frac{3C\mu_0^2 m_S m_p}{2\pi^2 r_{S-p}^8 M_p}\cot\theta_{S-p}\cos^2\theta_{S-p}$$

We see all orbits of ice particles at the end of their movement entering to magnetic equator plane. The solution $\Theta = \pi/2$ accounts for essentially planar structure of Saturn's rings and their location in the magnetic equator plane. The equation for azimuthal velocity of particles is:

$$\dot{\varphi}_{S-p} = \sqrt{GM_S/r_{S-p}^3 + 3C\mu_0^2 m_S m_p/(8\pi^2 r_{S-p}^8 M_p)}$$

We see the gravity force in the particle's orbit is counterbalanced by both the centrifugal force and the force of diamagnetic expulsion.

Then the model of spatially separated of dense packaging particles in the Saturn's rings we presented as uniformly magnetized spheres in a disk-shape structure consisting of the same spheres.

What we found:

- magnetization and magnetic moment of the disk-like structure is much higher than those of a single sphere due to the alignment of the multiple magnetic dipoles with the field.

- In addition, in a disk-shaped structure, the force of diamagnetic expulsion in the weak field region is stronger, and the magnetic well on the magnetic equator is deeper. This feature provides a sufficiently strong stability of the particles in the orbits and the entire ring system.

As can be seen from the above equations and from the modeling of the ring structure in the form of a disk-like structure, an important role in the appearance and stability of the ring system belongs to the axisymmetric magnetic field. Also we can see its role increases with decreasing particle size.

4. **Why spokes are located in the ring *B* only?**

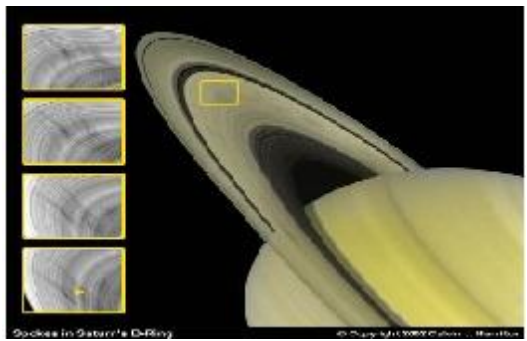

Fig. 2. Spokes in Saturn's *B*-ring. 2002 C.J. Hamilton

The spokes of the ring *B* are located almost radially. Their size is ~ $10^4$ km along the radius and about $10^3$ km along the orbit. The spectral radiation power has periodicity ~ 640.6±3.5 min which almost coincides with the period of rotation of the magnetic field of Saturn (639.4 min). There is correlation of maxima and minima of activity of spokes with spectral magnetic longitudes which is connected to presence or absence of the radiation of Saturn's Kilometric Radiation.

There are two important problems that are not yet answered. But we can do it with our theory:
1) why do spokes exist at all? 2) why are the spokes located only in the *B* ring, and there are no spokes in the other rings?

Important discovery was made by M. M. Hedman, P. D. Nicholson [12]. They found Saturn *B* ring has three times less matter than previously thought. P. Nicholson: "The best analogy is something like a fog over the meadow may seem less transparent and empty, than a water-filled pool, which has a much higher density than the fog". In other words *B* ring filled by small ice particles.

We can use our solution to explain the existence and features of the spokes in the *B* ring.

When the ice particles in the *B* ring enter the anomalous region of Saturn's magnetic field, the force of diamagnetic expulsion applied to the particles changes its value. Then the particles begin to change their orbit. For a huge number of particles, for an external observer, this process is represented as a turbulent cloud, stretched along the radius in the form of spokes. After passing through the anomaly, the particles return to their former orbit, and the appearance of the rings is restored.

The existence of the spokes becomes more understandable, since we found that the role of the axisymmetric magnetic field increases with decreasing particle size.

For the same reason, the spokes exist only in the *B* ring, because the *B* ring is made up of small particles

5. **What we can say about type of ice in the particles of Saturn's rings?**

Ice in the rings has existed for a long time. It is hardly possible to create such an ice in a laboratory on Earth to simulate its properties. Ice is a product of water. Recently was found unusual fact the deuterium to hydrogen isotopic ratio for Saturn's rings is the same as for the Earth [13]. It means that water in the Saturn's rings like water on the Earth. Now we have two possibilities to represent the properties of the ice particles of the rings. We can try to find some types of ice on Earth that can match the environmental parameters in Saturn's rings. Or we can also try to understand what properties ice particles can have from our proposed theory of the origin of Saturn's rings, if its data is consistent with the measurements of the Cassini probe. We know 17 types of ice on Earth. As it turned out, type XI ice has stable parameters at the temperature of the rings. Such ice was discovered in Antarctica, its age estimated as 100 years. It may originate of ordinary ice below -32,8°F. On the other hand, we found that it is possible to construct a theory of the origin of Saturn's rings if we assume that the ice in the particles of the protoplanetary cloud has diamagnetism.

Particles of Saturn's rings are known to consist mostly of ice [1]-[3] which is diamagnetic at not high pressures [14]. Below the atmospheric pressure, there is the low-temperature ice XI (Fig. 3) [15, 16]. The proton-ordered orthorhombic ice XI is stable below ~73 K. From an infrared image taken by the Cassini spacecraft, the temperature of Saturn's rings has been found to vary from 70 K to 110 K [17], and it is below 73 K over the entire area of the rings at equinox [18]. Thus, it is likely that the particles of Saturn's rings consist mostly of ice which is like ice XI.

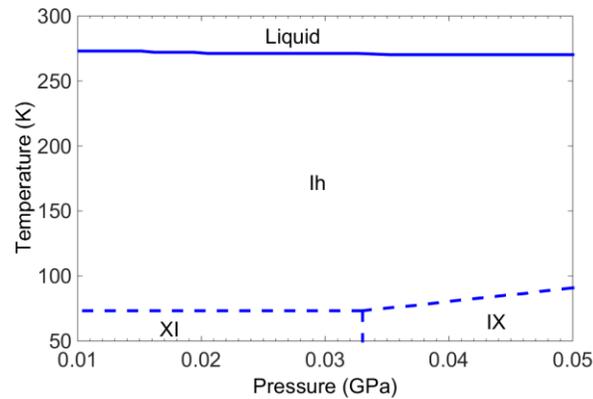

Fig. 3. Low-pressure part of the phase diagram of ice. Adapted from[14]. Below ~33 MPa, ice can form the high-temperature modification Ih and the low-temperature modification XI.

## 6. Conclusions

Here we have demonstrated how Saturn could create its rings by itself, with the additional action of its axisymmetric magnetic field once appeared. We started by solving the problem of the magnetization and, consequently, the potential energy of a spherical ice particle in the magnetic field of Saturn, the solution of which was later used to describe the dynamics of the particle in the superposition of the gravitational and magnetic fields of the planet.

We have shown that diamagnetic rings' particles may be responsible for the rings' genesis and stability. All diamagnetic materials in non-uniform magnetic fields are known to be expelled into the weak-field areas, and this feature provides a minimum of the potential energy of diamagnetic ice particles in the Saturn's magnetic equator plane. Taking this into account, the role of magnetism in the stabilization of planetary rings can be fairly universal. To understand the origin and evolution of Saturn's rings, and to shed light on their origin, magnetization and dynamics of ring particles in the magnetic field of the planet should be studied.

It follows from our solution, in the superposition of the spherically symmetric gravitational field and axially symmetric magnetic field, the orbits of particles finally fall on the magnetic equator plane and formatted rings structure of Saturn. Thus it also explains the essentially planar structure of the rings.

In the gravitational field only, the solution of the equations of motion yields the ratio of angular velocity components that turns out to be extremely unlikely, which speaks against the purely gravitational approach to Saturn's rings.

If the magnetic force is taken into account, the location of circular orbits of the particles in the magnetic equator plane consistently follows from the equation solution.

After appearance of magnetic field of Saturn collisions in motion of huge number of particles of protoplanetary cloud will compensate their azimuth-orbital movements and all their orbits should come to magnetic equator plane. As a result, all particles will be located on the Kepler's orbits, where there is a balance of gravity, centrifugal and force of diamagnetic expulsion.

The magnetic field in the plane of the disk of rings is inhomogeneous. Its lines will strive to pass through the region with the highest magnetic permeability, and diamagnetic particles will gather in the region with a low magnetic field density.

Density gradient flow of the magnetic field repels particles of each other and it also cleans the gaps within rings system and forms rigid thin structure of separated rings.

Each particle comes to the stable position at the magnetic equator of Saturn and preventing its own move up or down of it due to minimum magnetic energy at this position.

The horizontal displacement of the particles is prevented by the inhomogenity of the magnetic field along the radius, as is the case with iron particles around a magnet on a lab table.

Thus, it is likely that the particles of Saturn's rings consist mostly of ice which is like ice XI. Taking into account the diamagnetism of the ice particle ensures its more stable position in the plane of the Saturn's equator than if we take into account only the force of gravity.

The stability of diamagnetic ice particles in the magnetic well of the disk of Saturn's rings is stronger the more the number of particles in it, and the magnetic well itself will be deeper.

The spokes in the *B* ring have a clear explanation due to the diamagnetism of the ice particles of Saturn's rings. According to our theory, the effect of a magnetic field on a diamagnetic particle is stronger the smaller the particle size. This is most clearly seen in the B ring, because it consists of very small particles. Small particles of ice coming to the anomalous positions of the magnetic field of Saturn change their position, and we see it as spokes.

"NASA research reveals Saturn is losing its rings at "Worst-Case-Scenario" rate" [19]. At the same time such fatal pessimistic scenario is hard to believe because there is a backward process of "Icy tendrils going into Saturn ring traced to their source" [20] which is filling up rings by the particles coming from the geysers of satellite Enceladus due to magnetic coupling between Saturn and Enceladus [21]. Some day this situation may repeat again with another icy satellite started geyser activity next time because today there is a geological activity in space cause by the giant period cyclic change of the proton's energy [22].

Now it becomes clear how such a thin film of the huge diameter of Saturn's rings, consisting of ice particles, can hang in outer space.

We see that magnetic anisotropic accretion makes a significant contribution to the formation of Saturn's rings. This happens after the diamagnetic ice particles in the protoplanetary cloud begin to interact with Saturn's magnetic field after it appears.

It follows that the age of the rings may be close to the age of Saturn's magnetic field.

**References**

1. Saturn from Cassini-Huygens, M. Dougherty, L. Esposito, S. Krimigis (Eds.) (2009).
2. J. N. Cuzzi et al. Sci, 327, 1470 (2010).
3. L.W. Esposito. Annu. Rev. Earth Planet. Sci., 38, 383–410 (2010).
4. P. Estrada, R. Durisen, J. Cuzzi. AGU meeting. New Orleans, 2017, Dec. 12, paper 298112
5. A.I. Tsygan. Soviet Astronomy, 21 (4), 491-494 (1977).
6. A.M. Fridman, N.N. Gorkavyi. Physics of Planetary Rings (1999).
7. S. Charnoz et al. Icarus, 199 (2), 413-428 (2009).
8. R. M. Canup. Nature, 16, 468, 943–946 (2010).
9. V. Safronov. The evolution of the protoplanetary cloud and the formation of the Earth and planets, NASA (1972).
10. V.V. Tchernyi and S.V. Kapranov. ApJ, May 6, 894, 1 (2020). https://arxiv.org/abs/1907.07114, https://youtu.be/AI6AaMJoR4A
11. AAS Journal Author Series: Vladimir Tchernyi on 2020ApJ... 894...62T: https://youtu.be/La7RmcWGUTQ
12. M. M. Hedman, P. D. Nicholson. Icarus, 15, 279 (2016).
13. R. Clark et al, Icarus, 322 (2019)
14. F.E. Senftle, A. Thorpe, A. Nature, 194, 673-674 (1962)
15. R.J. Hemley. Annu. Rev. Phys. Chem., 51, 763-800 (2000).
16. V.F. Petrenko and R.W. Whitworth. Physics of Ice. Oxford, UK (2002).
17. C. Martinez. (2004), https://www.nasa.gov/mission_pages/cassini/media/cassini-090204.html
18. Spilker, L. et al. Icarus, 226, 316 (2013).
19. B. Steigerwald and N. Jones. (2018), https://solarsystem.nasa.gov/news/794/nasa-research-reveals-saturn-is-losing-its-rings-at-worst-case-scenario-rate
20. P. Dyches and S. Mullins. (2015), https://www.nasa.gov/jpl/cassini/icy-tendrils-reaching-into-saturn-ring-traced-to-their-source
21. Ghostly Fingers of Enceladus. (2014), https://www.jpl.nasa.gov/images/ghostly-fingers-of-enceladus
22. E.V. Chensky. IJAA, 3, 4, 438-463 (2013).

**Additional publications related to the topic:**

- V.V. Tchernyi et al. 52 LPSC Houston, March 15-19, 2021. [P651] POSTER SESSION: GIANT PLANETS, RINGS AND DYNAMICS — OH MY! https://www.hou.usra.edu/meetings/lpsc2021/pdf/1355.pdf

- V.V. Tchernyi et al. 43rd COSPAR Scientific Assembly, Sydney, Australia, 28 Jan. – 4 Feb. , 2021  https://www.youtube.com/watch?v=MBvpnihZFp4&feature=youtu.be

- V.V. Tchernyi et al. 52nd Annual DPS Meeting of AAS, Oct. 26-30, 2020. https://dps52-aas.ipostersessions.com/default.aspx?s=4A-8D-41-C6-A4-66-BC-91-29-69-56-0F-EE-C0-75-3D

- V.V. Tchernyi et al. The Eleventh Moscow International Solar System Symposium (11M-S3), Oct. 5-9, 2020. htttps://youtu.be/nHHS0lX6qtw

- V.V. Tchernyi et al. 235th AAS Meeting. Honolulu, HI, Jan. 4-8, 2020. Session 385 https://aas235-aas.ipostersessions.com/Default.aspx?s=aas_235_gallery

- V.V. Tchernyi and E.V. Chensky. IEEE GRSL, 2 (4), 445-446 (2005).

- V.V. Tchernyi and A.Yu. Pospelov. PIER, 52, 277 (2005).

- V.V. Tchernyi and E.V. Chensky. JEMWA, 19, 7 (2205).

- V.V. Tchernyi, A.Yu. Pospelov.  APSS, 307, 4, 347-356 (2007).

- V.V. Tchernyi. IJAA, 3, 4, 412-420 (2013).

- V.V. Tchernyi and A.Yu. Pospelov. IJAA, 8, 1, 104-120 (2018).